\documentclass[a4paper]{article}

\usepackage{INTERSPEECH2021}
\usepackage{url}

\title{NoiseVC: Towards High Quality Zero-Shot Voice Conversion}
\name{Shijun Wang$^1$, Damian Borth$^1$}
\address{
  $^1$University of St.Gallen, Switzerland}
\email{shijun.wang@unisg.ch, damian.borth@unisg.ch}

\begin{document}

\maketitle
\begin{abstract}
Voice conversion (VC) is a task that transforms voice from target audio to source without losing linguistic contents, it is challenging especially when source and target speakers are unseen during training (zero-shot VC). Previous approaches require a pre-trained model or linguistic data to do the zero-shot conversion. Meanwhile, VC models with Vector Quantization (VQ) or Instance Normalization (IN) are able to disentangle contents from audios and achieve successful conversions. However, disentanglement in these models highly relies on heavily constrained bottleneck layers, thus, the sound quality is drastically sacrificed. In this paper, we propose NoiseVC, an approach that can disentangle contents based on VQ and Contrastive Predictive Coding (CPC). Additionally, Noise Augmentation is performed to further enhance disentanglement capability. We conduct several experiments and demonstrate that NoiseVC has a strong disentanglement ability with a small sacrifice of quality.

\end{abstract}
\noindent\textbf{Index Terms}: voise conversion, vector quantization, contrastive predictive coding, augmentation

\section{Introduction}

Voice conversion (VC) is a task of converting the speaker characteristic information from one audio to another, while preserving the linguistic content. The idea of VC has been continuously studied and widely applied in many areas, e.g. privacy  protection, entertainment industry, \textit{etc.}

Most of previous works mainly focus on supervised VC \cite{Helander2010VoiceCU, Chen2014VoiceCU, Sun2015VoiceCU}, and achieve satisfactory performance. However, such methods need parallel training data, which is difficult to collect, and the alignment gaps between source and target utterances may cause the corruption of the voice conversion.

Recently, VC models without the requirement of parallel data gain more and more attention due to their effective utilization of non-parallel data. Models in \cite{Qian2011AFM, Trk2006RobustPT} attempt to find the optimal segments from non-parallel source-target pairs. ASR system is incorporated in \cite{Sun2016PersonalizedCT, Xie2016AKD} to help VC models synthesize speech from phoneme sequences. Generative models like variational autoencoder (VAE) \cite{Hsu2017VoiceCF, Kaneko2017ParallelDataFreeVC}, generative adversarial networks (GAN) \cite{Kaneko2018CycleGANVCNV, Kameoka2018StarGANVCNM} or Flow \cite{Serr2019BlowAS} are actively explored for the VC task as well. Despite successful conversion with non-parallel data, some of them produce fragile audios. Moreover, one major limitation is that they are not able to perform zero-shot conversions, i.e. synthesizing a voice of an unseen speaker.

One way to solve the zero-shot problem is recasting the VC task as a disentanglement problem, where linguistic contents and speaker information are disentangled from each other. Therefore, voice can be coverted by changing the speaker characteristic from the target speaker to the source, while keeping contents in the source. AutoVC \cite{Qian2019ZeroShotVS} applies a pre-trained speaker encoder and a vanilla auto-encoder with carefully designed bottleneck layers to force the encoder to only extract content information. Instance Normalization \cite{Ulyanov2017ImprovedTN} is applied in \cite{Chou2019OneshotVC, Chen2020AGAINVCAO} for zero-shot conversion, such idea has been used in the disentanglement task in computer vision domain. In \cite{Zhang2019JointTF, Zhang2020TransferLF}, a Text-to-Speech model is introduced to supervise the training of latent representations of a VC system. However, these models either lack adequate disentanglement ability or need assistance from text transcriptions, which is not desirable for low-resourced languages.

Recently, it has been observed that discrete latent code in auto-encoder based on Vector Quantization (VQ) \cite{Oord2017NeuralDR} are highly related to the phoneme \cite{Chorowski2019UnsupervisedSR}. Thus, VQ-based VC models \cite{Wu2020OneShotVC, Wu2020VQVCOV} can successfully disentangle contents from audios. However, effective disentanglement is gained from the massive sacrifice of the audio quality due to the constrain on the bottleneck. On the other side, works \cite{Oord2018RepresentationLW, Baevski2020vqwav2vecSL, wavaugment2020} prove that Contrastive Predictive Coding (CPC) can largely enhance the phoneme representation learning from speech audios. Such an idea has been implemented in VC model \cite{Niekerk2020VectorquantizedNN}, but they can only do many-to-many conversions. It is worth to mention that there is no guarantee that CPC can remove time-varying acoustic features, e.g. pitch \cite{hess2008pitch}. In this paper, we propose a zero-shot VC model based on VQ and contrastive learning. Additionally, we perform Noise Augmentation to further enhance disentanglement. We make the following three contributions:

\begin{itemize}
\item We propose a model, which can do zero-shot VC from unseen source speakers to unseen target speakers without harsh constrain on bottleneck layers.
\item We show that our VC model does not rely on parallel data, pre-trained models, or text transcripts, which is particularly desirable for low-resourced languages.
\item We demonstrate that our VC model can successfully disentangle linguistic contents and speaker characteristics without a drastic sacrifice of audio quality. 
\end{itemize}

\section{Approach}

In this section, we introduce three disentanglement methods used in our model: VQ, IN, CPC. Then, we explain the architecture of our proposed VC model NoiseVC.

\begin{figure*}[t]
  \centering
  \includegraphics[width=\linewidth]{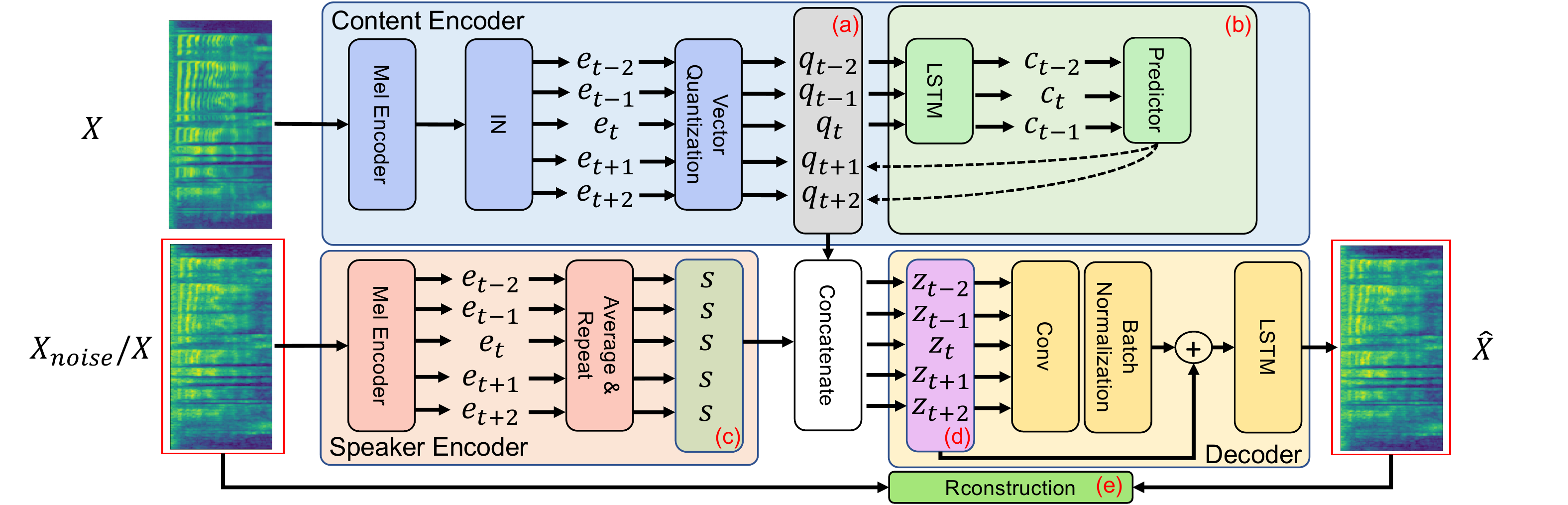}
  \caption{NoiseVC architecture. (a) Quantized sequence $\boldsymbol{Q}$ is considered the content embedding and the input of CPC module. (b) CPC module, it contains one LSTM layer and $k$ predictors, each $Pred^k$ needs to predict $q_{t+k}$. (c) Repeated sequence $\boldsymbol{S}$ is regarded as speaker embedding. (d) Content embedding and speaker embedding are concatenated into sequence $\boldsymbol{Z}$, which is fed to the decoder. (e) We alternatively feed noisy and original spectrograms into the speaker encoder, the reconstruction target is followed by this input.}
  \label{fig:architecture}
\end{figure*}

\subsection{Vector Quantization}

Vector Quantization \cite{Oord2017NeuralDR} maps continuous representations into discrete latent variables by using a shared codebook, while the decoder learns to do reconstruction with the quantized vectors. In audio domain, let $\boldsymbol{X} = \{x_1, x_2, ..., x_T\}$, where $T$ denotes time, be a sequence of acoustic features. An encoder is applied to transform $\boldsymbol{X}$ into a sequence of latent representation $\boldsymbol{E} = \{e_1, e_2, ..., e_T\}$, where each $e_t\in\mathbb{R}^D$ has dimension $D$. Each $e_t$ is quantized into a sequence of codes $\boldsymbol{Q} = \{q_1, q_2, ..., q_T\}$, where each $q_t$ is from a learnable discrete codebook $\mathcal{Q}^V$ with codebook size $V$. The quantization function can be written as:

\begin{equation}
  q_t = \arg\!\min_{q\in\mathcal{Q}^V}
  (\Vert{e_t-q}\Vert_2^2)
  \label{eq1}
\end{equation}

The quantization function takes the encoded $e_t$ and selects the closest $q$ from the codebook based on Euclidean distances.

A decoder is employed to reconstruct the input acoustic features with input $\boldsymbol{Q}$. The decoder output is denoted as $\boldsymbol{\hat{X}} = \{\hat{x_1}, \hat{x_2}, ..., \hat{x_t}\}$, and the reconstruction loss is:
\begin{equation}
  \mathcal{L}_{rec} = 
  \Vert{\boldsymbol{X}-\boldsymbol{\hat{X}}}\Vert_1^1 +
  \Vert{\boldsymbol{X}-\boldsymbol{\hat{X}}}\Vert_2^2
  \label{eq2}
\end{equation}

To update the codebook $\mathcal{Q}^V$ by gradient descent, we follow the objective from \cite{Oord2017NeuralDR}, so the total loss is:
\begin{equation}
  \mathcal{L}_{VQ} = L_{rec} + 
  \Vert{\mathrm{sg}[\boldsymbol{E}]-\boldsymbol{Q}}\Vert_2^2 + 
  \beta\Vert{\boldsymbol{E}-\mathrm{sg}[\boldsymbol{Q}]}\Vert_2^2
  \label{eq3}
\end{equation}
Where sg denotes the stopgradient operator. The second term means the codebook is updated by the encoder, and the last term prevents the encoder from growing arbitrarily. In all of our experiments, we set $\beta$ to 0.25.

\subsection{Instance Normalization}

Instance Normalization (IN) \cite{Ulyanov2017ImprovedTN} is a method originally used to do style transfer. Given input sequence $\boldsymbol{E}$, IN first computes the channel-wise mean $\mu$ and the channel-wise standard deviation $\sigma$. Then, the normalization is performed by:

\begin{equation}
  \textup{IN}(\boldsymbol{E}) = 
  \frac{\boldsymbol{E}-\mu(\boldsymbol{E})}
  {\sigma(\boldsymbol{E})}
  \label{eq:IN}
\end{equation}

Since $\mu$ and $\sigma$ are channel-wise and unrelated with time, they are regarded as representations with global information (speaker information). In the VC tasks, models in \cite{Chou2019OneshotVC, Chen2020AGAINVCAO} detach the global speaker information with IN method.

\subsection{Contrastive Predictive Coding}

Contrastive Predictive Coding (CPC) \cite{Oord2018RepresentationLW} is a self-supervised representation learning method. CPC loss is contrastive: positive future embeddings should be distinguished from negative future embeddings. In audio domain, this idea encourages the model to capture time-variant information, e.g. phoneme, while discarding time-invariant features \cite{Qian2019ZeroShotVS, Oord2018RepresentationLW, wavaugment2020, Schneider2019wav2vecUP}. 

In CPC, the sequence $\boldsymbol{E}$ from the encoder is passed to a recurrent network to produce context $\boldsymbol{C} = \{c_1, c_2, ..., c_T\}$. Then several predictors \textit{Pred}$^k$ ($0<k\leq K$) take $c_t$ and predict future representations $e_{t+k}$. With the contrastive loss, the prediction is achieved by minimizing the dot product between $e_t$ and correct future embedding while maximizing the dot product with negative embeddings $\mathcal{N}_{t,k}$ from the batch. The CPC loss can be given as:
\begin{equation}
  \mathcal{L}_{CPC} = \frac{1}{K}
  \sum\limits_{k=1}^{K}
  \textup{log}
  \frac{\textup{exp}( \textup{dot}( Pred^k(c_t) , e_{t+k}) )}
  {\sum_{n \in \mathcal{N}_{t,k}} \textup{exp} ( \textup{dot}( Pred^k(c_t), e_n ) )}
  \label{eq4}
\end{equation}

\begin{table*}[th]
  \caption{Speaker classification results on content embbeding $\boldsymbol{Q}$}
  \label{content probing}
  \centering
  \begin{tabular}{c | c c c c c}
    \toprule
    \cmidrule(r){1-2}
    Model               & $\alpha$      & Downsampling     & Codebook         & Speaker Accuracy(\%)              & L1 Reconstruction       \\
                        &               &    Factor        &  size / dimension     & (the lower the better)    &(the lower the better)     \\
    \midrule
    
    VQVC+               & -                  &   2; 2; 2             & 16/40; 32/20; 64/10            & 31.82; 15.91; 12.5        & 0.4602      \\
    NVC(IN+VQ)          & -                  &   -                   & 2048 / 512                     & 59.10                     & 0.3182      \\
    NVC(IN+VQ+CPC)      & -                  &   -                   & 2048 / 512                     & 30.68                     & 0.3870      \\
    NVC(IN+VQ+CPC)      & 0.3                &   -                   & 2048 / 512                    & 22.73                     & 0.3943      \\
    NVC(IN+VQ+CPC)      & 0.5                &   -                   & 2048 / 512                    & 21.59                     & 0.3945      \\
    NVC(IN+VQ+CPC)      & 0.7                &   -                   & 2048 / 512                     & 20.45                     & 0.3991      \\
    \bottomrule
  \end{tabular}
  
\end{table*}

\subsection{NoiseVC}

In this section, we describe our proposed model: NoiseVC. The architecture is illustrated in Figure \ref{fig:architecture}. The model contains two encoders, one to extract contents, the other is responsible for capturing speaker information. Additionally, we perform Noise Augmentation to enhance the disentanglement ability: two types of Mel-Spectrograms are fed as inputs. A decoder is applied to reconstruct one type of spectrogram with the concatenated outputs from two encoders. The objective of NoiseVC is the sum of VQ loss (Eq. \ref{eq3}) and CPC loss (Eq. \ref{eq4}).

\subsubsection{Content Encoder}

The content encoder is applied to extract time-variant contents. It contains four modules, mel encoder, IN, vector quantization, and CPC module. IN is applied for preliminary speaker information removing. VQ or CPC has been proved that the learned representations are very likely associated with phonemes. Also, inspired by previous work \cite{Niekerk2020VectorquantizedNN}, we use both methods together to efficiently produce content representations.

For each module, the mel encoder consists of 5 Convolutional layers with LeakyReLU activation. To avoid losing contents, we set the stride in all Convolutional layers to one and no polling operations are employed. After IN operation on the output of mel encoder, a sequence $\boldsymbol{E}$ is produced where each $e_t$ is 512D. Sequence $\boldsymbol{E}$ is then passed to the quantization module. To avoid content loss induced by a small codebook, a codebook contains 2048 codes with 512D is implemented. Such codebook setting is employed in all our models. The last CPC module takes quantized sequence $\boldsymbol{Q}$ as input, and produces context sequence $\boldsymbol{C}$ with an LSTM layer. To achieve contrastive loss, 34 predictors $Pred^k (0<k\leq 34)$ are built, each predictor needs to distinguish correct $q_{t+k}$ from 20 negative $q$ samples.

\subsubsection{Speaker Encoder}

The speaker encoder is built to extract time-invariant speaker characteristic features from spectrograms. Similar to the mel encoder in the content encoder, the speaker mel encoder includes 3 Convolutional layers with LeakyReLU activation function, the stride size is one as well. Since we expect the speaker encoder extract global time-invariant speaker features, the output sequence is then averaged and replicated to the length as long as the $\boldsymbol{Q}$ in the content encoder. The dimension is the same as content embedding, which is 512D.

\subsubsection{Decoder}

A decoder is implemented to do reconstruction. We concatenate the content embedding $\boldsymbol{Q}$ and repeated speaker embedding $\boldsymbol{S}$ as sequence $\boldsymbol{Z}$ to feed the decoder. So the input dimension is 1024. Decoder processes $\boldsymbol{Z}$ with 5 layers of Convolutional layers, while ReLU activation function, Batch Normalization and residual connection are included. The output of Convolutional layers is then passed to an LSTM network in order to reconstruct inputted Mel-Spectrograms.

\subsubsection{Noise Augmention} \label{Noise_Augmention}

I has been shown that VQ and CPC can successfully disentangle time-variant information while discarding time-invariant acoustic information. However, some acoustic features are time-varying as contents, e.g. pitch \cite{hess2008pitch}. Therefore, it is possible that the VQ function does not only extract contents, but also undesirably capture pitch especially when the codebook size is large. Additionally, there is no guarantee that the predictors in CPC can do right predictions only based on the preceding contents, since time-varying pitch can also be used for CPC.

To avoid such a situation and make the content encoder focus only on the content extraction, we perform Noise Augmentation. Specifically, we feed the model with two types of Mel-Spectrograms, one is the original spectrogram, the other is the same spectrogram but augmented by adding Gaussian Noise. The augmented spectrogram contains the same contents, but its time-varying pitch feature is changed by the addition of noise \cite{sukhostat2015comparative}. During the training of NoiseVC, we always feed the content encoder with original spectrograms, while we alternatively send two types of spectrograms to the speaker encoder. With this strategy, the content encoder is not able to always access the reconstruction target with a "real" pitch that needs to be reconstructed. Therefore, the content encoder is less likely to extract time-varying acoustic information and cares more about contents, since acoustic information from the content encoder is useless for the decoder to do reconstruction. The selection of reconstruction target is same as the input of the speaker encoder, when we input an augmented spectrogram to the speaker encoder, the decoder needs to reconstruct the augmented version, otherwise, the target is the original version. The choice of origin/augmentation is controlled by a hyper-parameter $\alpha$.

\section{Experiments}

In this section, we conduct experiments to verify the disentanglement performance and sound quality of NoiseVC. Audio demo can be found at \url{https://hsg-aiml.github.io/noisevc_demo/index.html}.

\subsection{Dataset and Preprocessing} \label{Dataset}

We conduct experiments on VCTK dataset \cite{Veaux2017CSTRVC}, which contains 110 English speakers, each speaker reads out about 400 sentences. After preprocessing, we select 108 speakers and split them into training and testing set. There are 88 speakers in the training set, which we denote them as seen speakers, the rest 20 speakers in the testing set are unseen speakers. In addition, for later probing evaluations, one audio sample from each seen speaker is randomly selected and mixed into the testing set. All audio files are downsampled from 48000Hz to 22050Hz, and silence is removed. Then, wave files are converted into 80-bin Mel-Spectrograms with 1024 STFT window size and 5.8 milliseconds hop size, small hop size is used because we expect our model to fully utilize the big codebook. To convert spectrograms back to waveform, a vocoder PwGAN \cite{Yamamoto2020ParallelWA} is pre-trained.

\subsection{Baseline Models}

We compare NoiseVC with two other State-of-the-Art VC models. AutoVC \cite{Qian2019ZeroShotVS} and VQVC+ \cite{Wu2020VQVCOV}. AutoVC uses speaker embedding provided by a pre-trained speaker encoder, such speaker embedding encourages its content encoder to focus on content extraction. VQVC+ is a U-net like VC model with hierarchical VQ layers. To make fair comparisons, these two models are retrained with the same dataset as NoiseVC.

\subsection{Content Encoder Study}

To verify the disentanglement ability of the content encoder, we train a speaker classifier on content embedding $\boldsymbol{Q}$ produced by the content encoder. The classifier consists of 3 Convolutional layers. Since AutoVC employs a pre-trained speaker embedding as part of its input, so it's not comparable here.

For successful VC, we expect our model to achieve low speaker classification accuracy on content embedding (good disentanglement) and low L1 reconstruction error (good sound quality) at the same time. The results are shown in Table \ref{content probing}. As we can see, VQVC+ achieves good disentanglement, but the reconstruction ability (audio sound quality) is heavily sacrificed. On the other side, our models, NVC(IN+VQ+CPC), are able to produce good-quality audios, while efficient disentanglement is preserved. It's also worth to mention that in VQVC+, apart from small codebooks, they also reduce their content sequence by a downsampling factor to further constrain the bottleneck layers. Contrarily, we implement a big codebook in NoiseVC without downsampling, to make sure all contents can be preserved.

Additionally, we test the disentanglement ability of CPC and Noise Augmentation. A model trained with only IN and VQ, NVC(IN+VQ), as shown in the table, still leads to imperfect disentanglement when a big codebook is applied. The model learns to copy inputs directly to the decoder, which reaches perfect but pointless reconstruction ability. If we apply the CPC to NoiseVC, the accuracy drastically drops from 59\% to 30\%. Moreover, the speaker classification accuracy can be further reduced with Noise Augmentation, as shown in the last three rows, we test three $\alpha$ values, and all of them bring better disentanglement ability, while maintaining similar reconstruction ability as the model without Noise Augmentation.

Our results demonstrate that CPC and Noise Augmentation help NoiseVC more focus on content extraction even when we use a big codebook that contains wide dimension codes. Meanwhile, NoiseVC can produce good quality audios due to its big codebook. In later experiments, we use NVC(IN+VQ+CPC) with 0.5 $\alpha$ as our default model.

\begin{figure}[t]
  \centering
  \includegraphics[width=\linewidth]{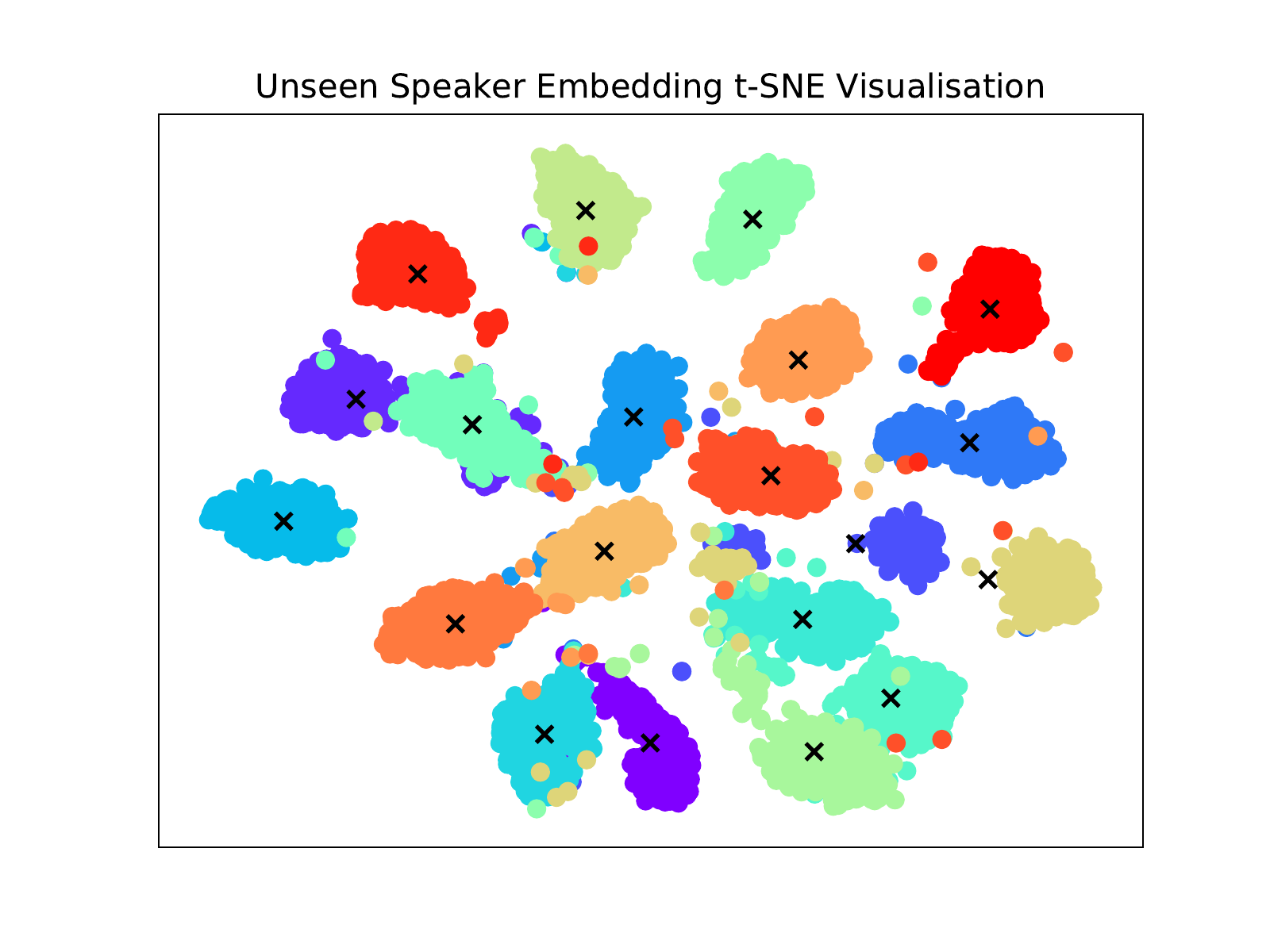}
  \caption{t-SNE visualisation on 20 unseen speakers embedding}
  \label{fig:speaker_emb}
\end{figure}

\subsection{Speaker Encoder Study}

To test the speaker encoder, a one-layer liner speaker classifier is trained to do speaker classification on the speaker embedding $\boldsymbol{S}$ produced from the speaker encoder. The speaker classifier achieves 98.9\% accuracy on evaluation dataset. Also, as shown in Figure \ref{fig:speaker_emb}, we plot speaker embedding of 20 unseen speakers with t-SNE \cite{Maaten2008VisualizingDU} method. Each unseen speaker has around 400 audio samples, the centroid of each speaker is labeled as a cross symbol. With the high speaker classification accuracy and clear 20 clusters in Figure \ref{fig:speaker_emb}, we demonstrate the speaker encoder does a successful speaker information extraction. 

\subsection{Subjective Evaluation}

We performed two subjective Mean Opinion Score (MOS) evaluation tests on Amazon Mechanical Turk (MTurk). Ground truth (oracle) samples are also scored to get the upper bound. For oracle samples, we first transform the wave files into spectrograms, then convert them into waveform with PwGAN.

The first MOS test is similarity test. In this test, the subjects are presented with pairs of utterances. Each pair has one converted sample, and one ground truth sample randomly selected from the target speaker. Subjects are asked to assign a score of 1-5 on the voice similarity: 5) Same, absolutely sure; 4) Same, sightly sure; 3) Not sure; 2) different, sightly sure; 1) different, absolutely sure. Each pair is assigned to 10 subjects. To generate converted samples, we randomly select 10 speakers from seen speakers and unseen speakers, respectively. We call them seen speaker set and unseen speaker set. For each speaker set, we then produce 10×9 = 90 conversions by generating a sample of each of the 10 speakers to each of the other 9 speakers. 

In the second test, we ask subjects to assign a score on the audio sound quality: 5) Excellent; 4) Good; 3) Fair; 2) Poor; 1) Bad. Same as in the similarity test, 90 samples are respectively produced from seen speaker set and unseen speaker set.

\begin{figure}[t]
  \centering
  \includegraphics[width=\linewidth]{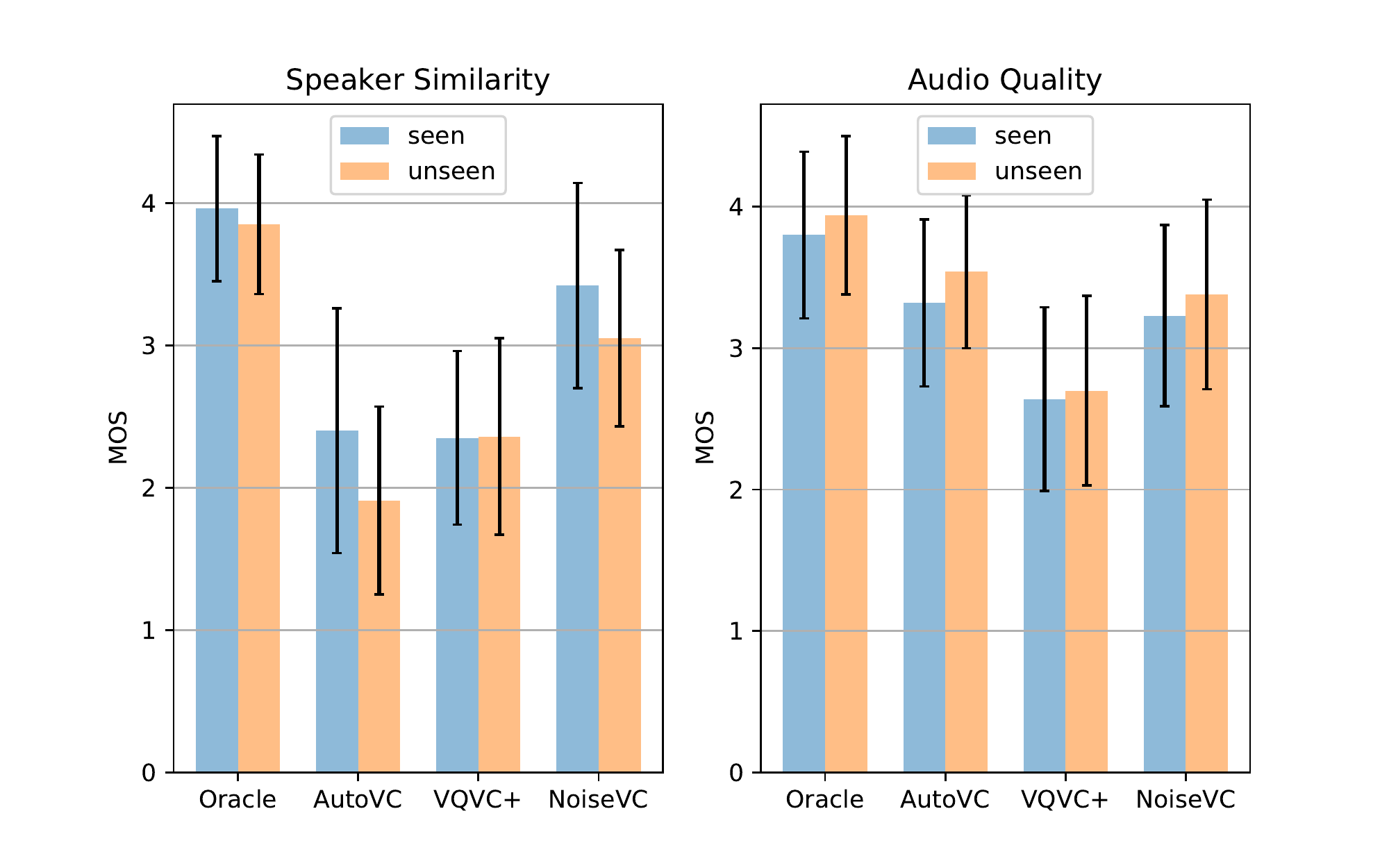}
  \caption{MOS scores on similarity and quality}
  \label{fig:mos}
\end{figure}

MOS scores are presented in Figure \ref{fig:mos}. For the speaker similarity test, NoiseVC outperforms both baseline models, and it can do a good job on zero-shot voice conversion. For the audio quality test, AutoVC reaches the highest score among the three models. However, according to the results in the speaker similarity test and some samples we listen to, the conversion voice is either the same as the source speaker, or the mix of source and target speakers (conversion fails). The reason could be caused by the small hop size we use when we do Mel transform, so AutoVC learns to copy the input to the decoder. On the other side, NoiseVC can produce audios with good quality due to the big codebook size. Moreover, it has the ability to do successful voice conversion at the same time. One thing we need to mention is that in quality test, the MOS scores of unseen are better than seen, such situation is caused by the lower quality of unseen ground truth data subset than seen subset, as showed in the MOS results for oracle where unseen score is better than seen.

\section{Conclusions}

In this work, we proposed NoiseVC, a zero-shot VC system that can disentangle contents from Mel-Spectrograms and generate high-quality conversion audios due to a big codebook. Experimental results showed that with CPC and Noise Augmentation, the content encoder achieves satisfactory disentanglement of the content information. Meanwhile, the speaker encoder is able to efficiently extract speaker characteristic information. Moreover, human evaluation strongly supports the effectiveness of utilized methods we implement in NoiseVC.

\bibliographystyle{IEEEtran}

\bibliography{main}


\end{document}